\documentclass[iop]{emulateapj-rtx4} % read by: ACROREAD
\shortauthors{Sekanina}
\shorttitle{2I/Borisov and Oort Cloud Comets}
\slugcomment{Version \today }

\begin{document}
\title{Are 2I/Borisov and Oort Cloud Comets Alike?}
%\title{WHAT TO EXPECT FROM THE FIRST ACTIVE INTERSTELLAR COMET?}
%
\author{Zdenek Sekanina}
\affil{Jet Propulsion Laboratory, California Institute of Technology,
  4800 Oak Grove Drive, Pasadena, CA 91109, U.S.A.}
\email{Zdenek.Sekanina@jpl.nasa.gov.{\vspace{-0.27cm}}}

\begin{abstract}
The interstellar comet 2I/Borisov bears a strong resemblance to Oort
Cloud comets, judging from its appearance in images taken over the first
six weeks of observation.  To test the proposed affinity~in~more
diagnostic terms, 2I is compared to Oort Cloud comets of
similar perihelion distance, near 2~AU.  Eight such objects are
identified among the cataloged comets whose orbits have been
determined with high accuracy.  This work focuses on three particular
characteristics:\ the light curve, the geometry of the dust tail, and
the dust parameter {\it Af}$\rho$.  Unlike Oort Cloud comets with
perihelia beyond the snow~line, Oort Cloud comets with perihelia near 2~AU show
strong evidence of the original halo of slowly accelerating, millimeter-sized
and larger icy-dust grains only in early tail observations.  The dust tail
in later images is primarily the product of subsequent,
water-sublimation driven activity nearer perihelion but not
of activity just preceding observation, which suggests the absence of
microscopic-dust ejecta.  Comet 2I fits, in broad terms, the properties
of the Oort Cloud comets~with perihelia near 2~AU and of
fairly low activity.  Future tests of the preliminary conclusions are
proposed.
\end{abstract}


\section{Introduction}
The era of interstellar comets is upon us, having had so far the flavor of a
bizarre story.  Less than two years~after the hectic weeks with a weird
cometary cocoon designated 1I/`Oumuamua, whose nature is still shrouded in
mystery, Gennady Borisov, a Russian {\it amateur\/} astronomer, has pulled the
proverbial rabbit out of a hat by discovering, on 2019 August 30, the {\it
first active interstellar comet\/}, of apparent magnitude 18, with a {\it 65-cm
telescope} (!) of his own design (Green 2019, O'Callaghan 2019).  Borisov's find
is a stunning achievement by any measure, especially in the light of the
 post-`Oumuamua nearly universal consensus that instrumentation more powerful
than currently available is needed to successfully search for further interstellar
objects.  While this perception is prudent in follow-up studies, Borisov's
discovery demonstrates that ever bigger telescopes are~by~no means indispensable
in meeting the basic objective.

While the comet's interstellar origin had~been~flagged by a hyperbolic-motion
search code on September~8 (Guzik et al.\ 2019), the official
designation 2I/Borisov was inexplicably delayed until September~24, or 25~days
after discovery.  To assign the designation 1I to `Oumuamua took only 19~days,
even though groundbreaking and time-consuming IAU negotiations were then involved.

The new comet is bound to trigger revised estimates of the population of
interstellar comets.  On the one hand, one should be wary of the law of
small numbers, on the other hand it is true that discoveries of comets
of apparent magnitude around 18 have been fairly common for close to 70~years,
essentially since the time the 122-cm Schmidt telescope became operational
at Palomar in late 1948 (e.g., Harrington 1952).  Records show that
comet C/1954~M1 was at the time of discovery of magnitude 19 (e.g., Porter  
1955).  It thus took at least 65~years to discover the first interstellar
comet as faint as 2I/Borisov at a fairly large geocentric distance.

\section{The Orbit Eccentricity}
Although the orbit eccentricity of 2I/Borisov is much higher than that
of 1I/`Oumuamua, both values are approximately what one expects, if either
object arrived from a star system whose velocity relative to the Sun is near
a statistically averaged value determined by modeling the kinematics of
main-sequence stars and the Sun's motion with respect to the Local Standard
of Rest (LSR).  For the averaged velocity $V_{\rm rel}$ of an interstellar
object upon its approach to the Sun one finds
\begin{equation}
\langle V_{\rm rel}^2 \rangle = V_{\rm sun}^2 \!+ \langle \sigma_{\rm V}^2 \rangle,
\end{equation}
where $V_{\rm sun}$ is the Sun's velocity with respect to the LSR and
$\sigma_{\rm V}$ is the mean dispersion of star velocities, which according
to Dehnen \& Binney (1998) is a function of the $B \!-\! V$~color, increasing
from blue to red stars.  These authors find a color independent \mbox{$V_{\rm
sun} = 13.4$ km s$^{-1}$}{\vspace{-0.04cm}} and from their results it follows that
\mbox{$\langle \sigma_{\rm V}^2 \rangle^{\frac{1}{2}} = 28.3$ km s$^{-1}$} for
stars{\vspace{-0.04cm}} in the $B \!-\! V$ range of +0.14 to +0.47~mag.~Equation~(1)
yields \mbox{$\langle V_{\rm rel}^2 \rangle^{\frac{1}{2}} = 31.3$ km s$^{-1}$} and
the eccentricity $e$ is related to the perihelion distance $q$ (in units~of AU) by
\begin{equation}
e = 1 + \frac{\langle V_{\rm rel}^2 \rangle}{V_{\rm earth}^2} \, q ,
\end{equation}
where \mbox{$V_{\rm earth} = 29.78$ km s$^{-1}$} is the Earth's mean orbital
velocity.  For `Oumuamua \mbox{$q = 0.255$ AU} and the predicted \mbox{$e =
1.28$} is close to the actual value of 1.20; for Borisov's comet \mbox{$q =
2.01$ AU} and the predicted \mbox{$e = 3.22$} is not too far from the actual
value of 3.36.\footnote{Comparison made with the elements in MPEC
2019-T116.{\vspace{-0.15cm}}}

On the other hand, the eccentricity effect on orbital velocities
of the two objects turns out to be dramatic.  At the perihelion and discovery
times the parabolic limits were exceeded, respectively, by 8~percent and
21~percent for 1I/`Oumuamua, but by 48~percent~and 66 percent in the case
of 2I/Borisov.

\begin{table*}[t]
\vspace{-4.2cm}
\hspace{-0.5cm}
\centerline{
\scalebox{1}{
\includegraphics{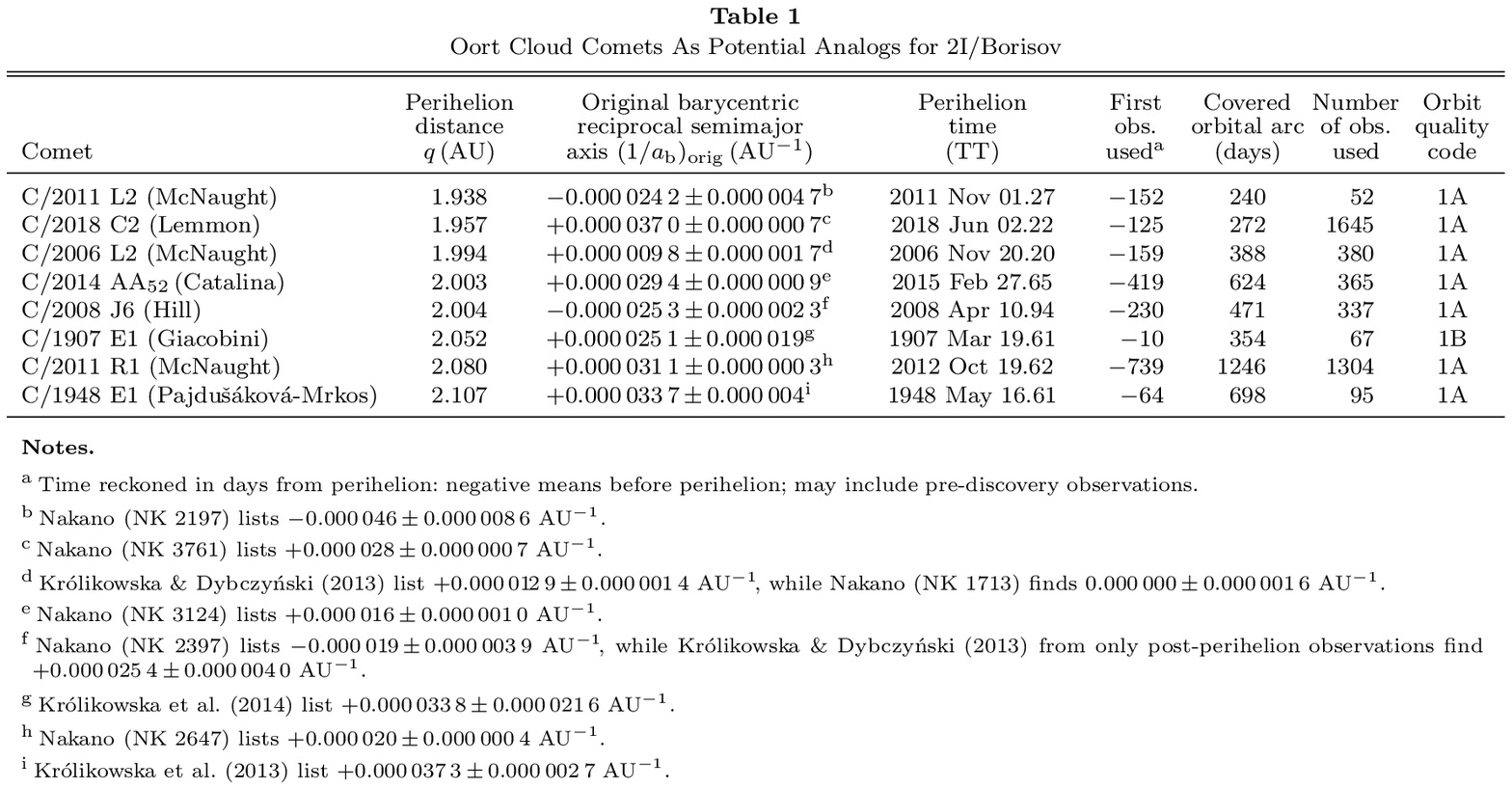}}}
\vspace{-15.31cm}
\end{table*}

\section{1I/`Oumuamua Contra 2I/Borisov:\ Traits Shared with Oort Cloud Comets}
Investigations of the physical and compositional properties of 2I/Borisov have
already begun (e.g., Guzik et al.\ 2019, Fitzsimmons et al.\ 2019, Yang et
al.\ 2019;~de Le\'on et al.\ 2019; Dybczy\'nski et al.\ 2019; Jewitt \& Luu
2019; Kareta et al.\ 2019) and will surely continue.  Cursory inspection of the
appearance of the comet suggests its {\it striking similarity to Oort Cloud
comets\/}, unlike 1I/`Oumuamua, which has often been described~as~an~object
unlike any other ever observed.  Yet, `Oumuamua could be a piece
of debris of a dwarf interstellar comet, which, because of its small perihelion
distance, had disintegrated near perihelion (Sekanina 2019a).  Intrinsically
faint, dust-poor Oort Cloud comets with perihelia closer to the Sun than
$\sim$0.5~AU are known to do just that (Sekanina 2019b).  In the process of
disintegration, the properties of the parent body altered to the extent that
`Oumuamua as a fragment could not be expected to provide information on the
body that had arrived from interstellar space.  2I/Borisov is expected to be
largely immune to such a mishap because of its distant orbit.

Most members of the Oort Cloud population have their perihelion distances beyond
the snow line and their impressive preperihelion activity at large heliocentric
distances is being driven by processes other than the sublimation of water ice.
Referring to three such objects --- C/1954~O2 (Baade), C/1954~Y1 (Haro-Chavira),
and C/1956~F1 (Wirtanen) --- Roemer (1962) described their appearance by remarking
that a featureless, rectilinear tail emanates from a clearly bounded envelope
around the nuclear condensation.  The tail is very nearly parallel-sided, its
width essentially unchanged and equal to the diameter of the head up to the
point of extremity.  This applies equally after perihelion, when dust tails of
other comets have a tendency to broaden considerably.

Osterbrock (1958) noted that the tails of comets Baade and Haro-Chavira were
oriented about midway between the antisolar direction and the orbit behind the
nucleus.  This puzzle was eventually settled by Sekanina's (1973, 1975) proposal
of the conservation-of-momentum law's effect on {\it sizable dust grains\/}
released at {\it very low velocities\/} from the nucleus {\it far from the Sun\/}
(up to 10~AU or more) on the way to perihelion.  The problem of activity of Oort
Cloud (and some other) distant comets long before perihelion was recently
revisited by Meech et al.\ (2009).

If active interstellar comets, such as 2I/Borisov, and Oort Cloud comets are
alike, their similarity should extend far beyond their mere appearance.  If so,
it is essential
that the dependence of cometary activity on heliocentric distance be recognized
in undertaken tests of diagnostic physical signatures.  Accordingly, the behavior
of 2I/Borisov should be compared with Oort Cloud comets of similar perihelion
distance.

\section{Potential Analogs to 2I/Borisov}
Since the orbital computations showed that the perihelion distance of 2I/Borisov
equals almost exactly 2.0~AU, I searched for cataloged long-period comets whose
orbits match this condition to within $\pm$0.1~AU.  I found a total of 26~objects
since 1890, of which nine were observed poorly enough that only a parabolic
orbit could be computed.  Of the 17~remaining entries only eight were {\it bona
fide\/} Oort Cloud comets, with an original barycentric reciprocal semimajor axis,
{\vspace{-0.04cm}}$(1/a_{\rm b})_{\rm orig}$, not exceeding +0.000\,050~AU$^{-1}$,
i.e., with the aphelion distance not below 40\,000~AU; they are listed in Table~1
in the order of increasing perihelion distance.\footnote{Two comets, C/1890 F1
{\vspace{-0.04cm}}(Brooks) and C/1993 A1 (Mueller), have $(1/a_{\rm b})_{\rm
orig}$ that is larger than +0.000\,050 AU$^{-1}$ but{\vspace{-0.04cm}} smaller
than +0.000\,100 AU$^{-1}$, i.e., aphelia between 20\,000 and 40\,000~AU.}
The values of $(1/a_{\rm b})_{\rm orig}$ are taken from the website of the
{\it Minor Planet Center\/}\footnote{See {\tt
https://minorplanetcenter.net/db\_search}.} (MPC), but independent values
from two other sources, the Warsaw group and S.~Nakano, are presented in the
footnotes to Table~1.  The orbit quality in the last column is described by
the classification scale introduced by Marsden et al.\ (1978).  An important
point is that none of the tabulated comets required inclusion of nongravitational
terms into the equations of orbital motion.  However, it is noted
that for two of the eight comets \mbox{$(1/a_{\rm b})_{\rm orig} < 0$}, a
potential signature of small nongravitational effects that have been unaccounted
for (Marsden et al.\ 1978).

The case of C/2008 J6 is particularly suspicious because Kr\'olikowska \&
Dybczy\'nski (2013) --- unlike either the MPC or Nakano --- obtained a positive
value~of $(1/a_{\rm b})_{\rm orig}$ by skipping the pre-discovery single-night
observations made by the Siding Spring Survey more than seven months before
perihelion (see below).
% and obtained a positive value of $(1/a_{\rm b})_{\rm orig}$.
On the other hand, they attempted a nongravitational orbital solution for
comet C/2006~L2 and did not find it at~all~helpful.  The Oort Cloud analogs
thus suggest that a measur\-able {\it nongravitational acceleration in the orbital
motion~of 2I/Borisov should not be ruled out\/} altogether, yet its~detection
may be difficult and the magnitude certainly much lower than for
1I/`Oumuamua.

In the following, information is presented for each of the eight Oort Cloud
analogs of 2I/Borisov, arranged chronologically. \\

\noindent {\it Comet C/1907 E1 (Giacobini)\,\/}\footnote{Old designation 1907a =
1907 I.} \\[-0.15cm]

Discovered on 1907 March 9 (Bassot 1907), only one week before perihelion, this
comet was poorly observed for brightness, in part because it was moving toward
the Sun in the sky and became lost in twilight.  The peak apparent magnitude of
11 was consistently reported by at least four independent observers within 72~hours
of the discovery, but it is not possible to assess the degree to which do these
estimates represent the comet's total brightness.

Using the ephemeris by Weiss (1907a, 1907b), Wolf (1907) recovered the comet on
December~4.  Much fainter than at the time of discovery, it was then observed
until 1908 February 26, which allowed the computation of an orbit of high enough
quality to recognize the object as a member of the Oort Cloud (Marsden et al.\
1978). \\

\noindent {\it Comet C/1948 E1 (Pajdu\v{s}\'akov\'a-Mrkos)}\,\footnote{Old
designation 1948d = 1948 V.{\vspace{-0.2cm}}} \\[-0.15cm]

The discovery of this 10th magnitude object was made first by L.\
Pajdu\v{s}\'akov\'a and then, dispelling her subsequent doubts about the find
by getting a photographic image, A.\ Mrkos on 1948 March 13 (Merton 1949), more
than nine weeks before perihelion.  Van Biesbroeck (1950) later detected
the comet's images on photographic plates taken on February~15 and March~5,
extending the preperihelion orbital arc covered by observations to 91~days.

A systematic series of 132 visual-brightness estimates was secured by Beyer
(1950), who also provided data sets on the orientation and length of the comet's
dust tail (Section~6).  Comparison shows that the magnitudes by Beyer and Van
Biesbroeck were entirely compatible,~so that the February~15 pre-discovery data
point~can~readily be linked with the rest of the observations. \\

\noindent {\it Comet C/2006 L2 (McNaught)} \\[-0.15cm]

A product of the Siding Spring Survey, this comet was discovered by McNaught
(2006) on 2006 June 14, more than five months before perihelion when it was of
magnitude 13.7.  An extensive set of CCD photometric observations, reported as
total magnitudes, was provided by K.\ Kadota (observatory code 349; footnote~3)
between August 2006 and April 2007. \\

\noindent {\it Comet C/2008 J6 (Hill)} \\[-0.15cm]

This object was not discovered until five weeks~after perihelion (Hill 2008),
when it was of magnitude~15.5~and had a clear nuclear condensation.  The only
preperihelion observations are CCD images found years later on four exposures
taken nearly nine months before discovery~on a single night by McNaught
(2013).~Because~of~the~post-perihelion discovery, information on this comet
is not used in the following. \\

\noindent {\it Comet C/2011 L2 (McNaught)} \\[-0.15cm]

Another catch by McNaught (2011a), of magnitude 18.1, was secured on 2011 June 2,
about five months before perihelion.  The comet was observed over a period of less
than three months, followed by a five-months gap of no reported observation.  The
object was then at high southern declinations, yet staying more than 50$^\circ$
from the Sun.  Resuming only in mid-January 2012, the observations terminated
before the end of that month.  Because of the gap, the comet is not explored
below. \\

\noindent {\it Comet C/2011 R1 (McNaught)} \\[-0.15cm]

Discovered on 2011 September 3, more than a year before perihelion, with a strongly
condensed coma and of magnitude 16.5 (McNaught 2011b), this comet was later
detected on 12~exposures obtained at the Catalina Sky Survey on 2010 October~10
and November~6 and 30 (Hill et al.\ 2012) and on three exposures taken at the
ISON-NM Observatory, Mayhill, also on October~10 (Elenin 2012), thus extending the
preperihelion orbital arc under observation to slightly more than two years.
Since the comet was last detected in mid-April 2014, the observations cover
nearly 3.5~years, clearly the record among the objects in Table~1.  Images from
2013 show a long, narrow tail, as the Earth was approaching its transit across
the comet's orbital plane. \\

\noindent {\it Comet C/2014 AA$_{\mbox{\footnotesize \it 52}}$
 (Catalina)} \\[-0.15cm]

When discovered by Kowalski (2014) in Catalina Sky Survey images from 2014 January
11, about seven weeks before perihelion, this object --- slightly brighter than
magnitude 20 --- had a stellar appearance and was classified as an asteroid.
However, when imaged near perihelion, on February 24, by Bolin et al.\ (2014)
of the Pan-STARRS Project at Haleakala, the object had a strongly asymmetric
appearance (also confirmed at Siding Spring) and was reclassified as a comet
(Williams 2014).  This report also listed two pre-discovery astrometric
observations from Haleakala.  A slender tail is prominently seen in images
taken months before perihelion.

Even though the comet was under observation over 21~months, until late-September
of 2015, there were~two major gaps:\ the first extended from late May to late
September 2014, the second from late February to late July 2015; both were
centered on times of small solar elongations. \\

\noindent {\it Comet C/2018 C2 (Lemmon)} \\[-0.15cm]

An apparently asteroidal object of magnitude 20, discovered by the Mount Lemmon
Survey (2018) on 2018 February~5, four months before perihelion, was found to
be marginally extended with a faint tail in CCD images taken at Mauna Kea on
March 22 (Micheli 2018).  Additional Lemmon images were soon found from January
28 (Gibbs et al.\ 2018).

This appears to be the intrinsically faintest of the eight comets in Table~1.
The observations terminated while the object was still at least three magnitudes
brighter than at discovery, but its solar elongation was dropping rapidly into
twilight. \\

Since no data on the chemical composition appear to exist on these
comets that could be used diagnostically to examine the degree of
correspondence in the behavior of 2I/Borisov~and~the~Oort
Cloud comets, I focus in the following on three fundamental physical
characteristics, namely:\\[-0.25cm]

(1) the overall light curve, including the dependence of the preperihelion
intrinsic brightness on heliocentric distance, and potential short-term
variations;\\[-0.25cm]

(2) imaging of a narrow, featureless dust tail, its~orien\-tation and
projected length as a function of time; and \\[-0.25cm]

(3) the {\it Af}$\rho$ parameter, a dust production proxy introduced by
A'Hearn at al.\ (1984), as a function of heliocentric distance along
the preperihelion arc of the orbit.

\section{The Light Curves}
A temporal variation of a comet's total brightness normalized to a unit
distance from the Earth provides important information on the object's
activity.  It not only describes the systematic rate of brightening
before perihelion and fading after perihelion, but also shows the
position of the brightness peak relative to perihelion and discloses
instances of flare-ups, whether sudden outbursts or more gradual surges
as well as of unexpected dips in brightness.  The rate of variation with
heliocentric distance $r$ is diagnostic of the comet's physical state; for
example, an $r^{-2}$ trend for a dust-rich comet is characteristic of a
constant cross-sectional area of grains in the atmosphere.

The enormous temporal span of the comets in Table~1, more than a
century, entails a problem for maintaining uniformity in the presentation
of the light curves because of the light-detector evolution over this
period of time.  Chronologically, the first two comets --- C/1907~E1 and
C/1948~E1 --- come from the era of visual and photographic observations,
while the remaining comets from the era of CCD observations, even though
traditional inspection of comets by visual methods has continued to the
present time.

To meaningfully relate the light curves derived by different techniques,
it is essential to establish standards allowing comparison of photometric
systems of individual observers.  There are at least three issues involved:\
(i)~the relationship between the total magnitudes derived visually and
from CCD images; (ii)~the major difference between the photometric
behavior of the nuclear condensation and the comet as a whole; and (iii)~the
usage of a variety of scanning apertures and color filters in CCD
imaging.  An additional problem is that of correcting for a phase effect,
which is being taken care of in this paper by applying the Marcus (2007)
standard law for dust-rich comets.  Phase angles are limited to $<\,$30$^\circ$.

As far as I know, only CCD magnitudes have so far been reported for 2I/Borisov.
Accordingly, I collected this type of photometric data for the selected Oort
Cloud analogs as well.  As described in some detail elsewhere (Sekanina 2017),
complications could be mitigated by restricting the CCD data set to total
magnitudes only and by introducing a system of appropriate corrections for
individual observers to convert their data to a common photometric scale.
Ultimately, the scaled total CCD magnitudes of the analog comets have been
converted to total visual magnitudes (normally obtained by estimating bright
comets with the naked eye, binoculars, or other small instruments), using
comparisons of a few overlapping data obtained by either method for C/2011~R1
when it was at its brightest.  These have shown that --- to the extent allowed
by the uncertainties involved --- the total brightness is visually about
1.5~mag higher than the total CCD magnitudes indicate.

Given that light curves with the preperihelion branch missing or with
gaps extending over long periods of time are deemed unsuitable for this
type of investigation, only four comets from Table~1 have qualified as
potential 2I light-curve analogs of interest --- C/1948~E1, C/2006~L2,
C/2011~R1, and C/2018~C2.  The light curves, in which the estimated total
visual brightness is represented by the magnitude $H_\Delta$,{\vspace{-0.01cm}}
normalized to a geocentric distance of 1~AU and to a zero phase angle, are
plotted in Figure~1.  They provide information on the nature of these objects
to be confronted with the light curve of 2I/Borisov when available.  First,
the scaling errors notwithstanding, enormous differences clearly exist in the
degree of activity among the four comets, equivalent at perihelion to a factor
of $\sim$300 in brightness.  Second, the two comets for which the data are
about evenly distributed before and after perihelion --- C/2006~L2 and
especially C/2011~R1 --- suggest that the post-perihelion branch of the light
curve is steeper, a well-known property of Oort Cloud comets (e.g., Whipple
1978).  An inflection or a possible minor dip before perihelion are apparent on
either of them.  The pre-discovery observations of C/2011~R1, made about two
years prior to perihelion, are well outside the boundary of the plot in Figure~1,
but they closely conform to the light-curve fit.  Evidence for C/2018~C2 is
inconclusive because the post-perihelion branch of its light curve is
cut short by the comet's disappearance in twilight, while C/1948~E1 appears
to have experienced a brightness surge (but not an outburst) in the weeks just
before perihelion.  And third, with the exception of C/1948 E1, whose light
curve near its maximum is distorted by the surge, {\hspace{0.4pt}}the
{\hspace{0.4pt}}comets {\hspace{0.4pt}}achieve {\hspace{0.4pt}}peak
{\nopagebreak}brightness

\begin{figure*}
\vspace{-3.65cm}
\hspace{-0.2cm}
\centerline{
\scalebox{0.69}{
\includegraphics{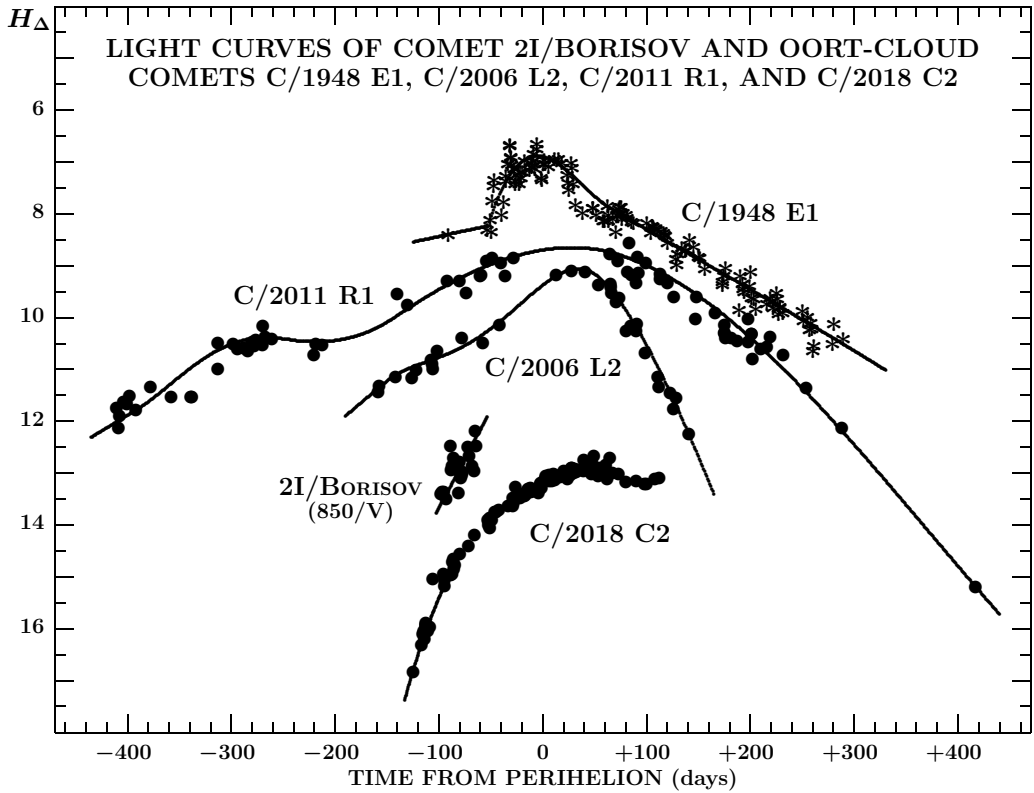}}}
\vspace{-6.32cm}
\caption{Light curves of four Oort Cloud comets with perihelia near 2~AU
and a version of 2I/Borisov's light curve.  For C/2006~L2 and C/2011 R1, the
total CCD magnitudes normalized to a unit geocentric distance, corrected for
the phase effect using the Marcus (2007) standard law for dust-rich comets,
and converted approximately to the visual photometric scale, $H_\Delta$, are
plotted against time, reckoned from perihelion time.  For C/2018~C2 there was
no difference between total and nuclear CCD magnitudes.  For C/1948~E1 the
plotted data are Beyer's (1950) uncorrected visual magnitudes.  Different
symbols used for C/1948~E1 prevent confusion with the overlapping data for
C/2011~R1.  The pre-discovery observations of C/2011~R1 are way outside the
plot, but fit the light curve. The only points plotted for 2I/Borisov are the
corrected data from code 850 (with use of a $V$ filter), but because of the
comet's excessively high orbital velocity, the relation between the brightness
and time does not properly reflect the position of the interstellar comet
relative to the Oort Cloud analogs; a plot against heliocentric distance in
Figure~2 is more representative.{\vspace{0.5cm}}}
\end{figure*}

{\noindent}some 6--7 weeks after perihelion.

For 2I/Borisov, the set of plotted data presents daily averages of
the code 850 magnitudes, obtained with a $V$ filter and reported on
the Minor Planet Electronic Circulars (MPEC).  As of early October,
this has been the most extensive set of data available from a single
observing site (20~points).  Each is the mean of several, usually
three, individually reported entries, some of which turned out to be
remarkably consistent with one another, whereas others showed fairly
large scatter.  After some experimentation, the factor used to convert
the averaged magnitudes to the adopted standard system of total visual
magnitudes was found to be about $-$2.1~mag.  Cursory inspection shows that
compared to the light curves of the Oort Cloud analogs the data are
surprisingly noisy.  The location of 2I in the plot is closer
to the lower end of the intrinsic-brightness range, but this may not
necessarily be the case.  Two important points are in order: (i)~the
CCD magnitude determinations vary from observer
to observer, so the data presented here are largely unsupported; and
(ii)~the plot of intrinsic brightness against time does not do the
comparison with the Oort Cloud comets justice because of 2I's much
higher orbital speed.

To remedy the problem pointed out in (ii), Figure 2 plots the {\it
preperihelion\/} normalized magnitudes $H_\Delta$ as a function of
heliocentric distance and fits the data by power laws, neglecting the minor
dip showed in Figure~1 by C/2011~R1.  Overall, the slopes are much steeper
{\vspace{-0.04cm}}than $r^{-2}$, indicating that the comets were unquestionably
active.  However, with the exception of C/1948~E1, the steepness
of the light curves near perihelion decreases with increasing intrinsic
brightness.  The pre-discovery observations of C/2011~R1
two years before perihelion (at distances of up to 7.5~AU from the Sun)
are now plotted at the extreme right of the figure.  The brightness of this
comet at heliocentric distances below 4~AU varies as $r^{-2.3}$, a rate
that jumps up rather abruptly to $r^{-6}$ beyond 4~AU.  A similar behavior
is revealed by C/2018~C2, except that the numbers{\vspace{-0.04cm}}
are very different:\ the comet's brightness varies as $r^{-11}$ below 2.3~AU
from the Sun, but as $r^{-22}$ farther from the Sun.  Interestingly,
the total and nuclear CCD magnitudes reported for this comet were
nearly identical, suggesting that it consisted of practically nothing
but the nuclear condensation.  Comet C/2006~L2 was not discovered early
enough to exhibit the break in the slope, but the data show that the
near-perihelion law $r^{-3.7}$ applied to a distance of at least 2.8~AU.
The extremely steep slope for C/1948~E1 is of course meaningless, merely
fitting the brightness surge over a period of time during which the
comet's heliocentric distance varied only very insignificantly.

\begin{figure*}[t]
\vspace{-1.6cm}
\hspace{-0.32cm}
\centerline{
\scalebox{0.69}{
\includegraphics{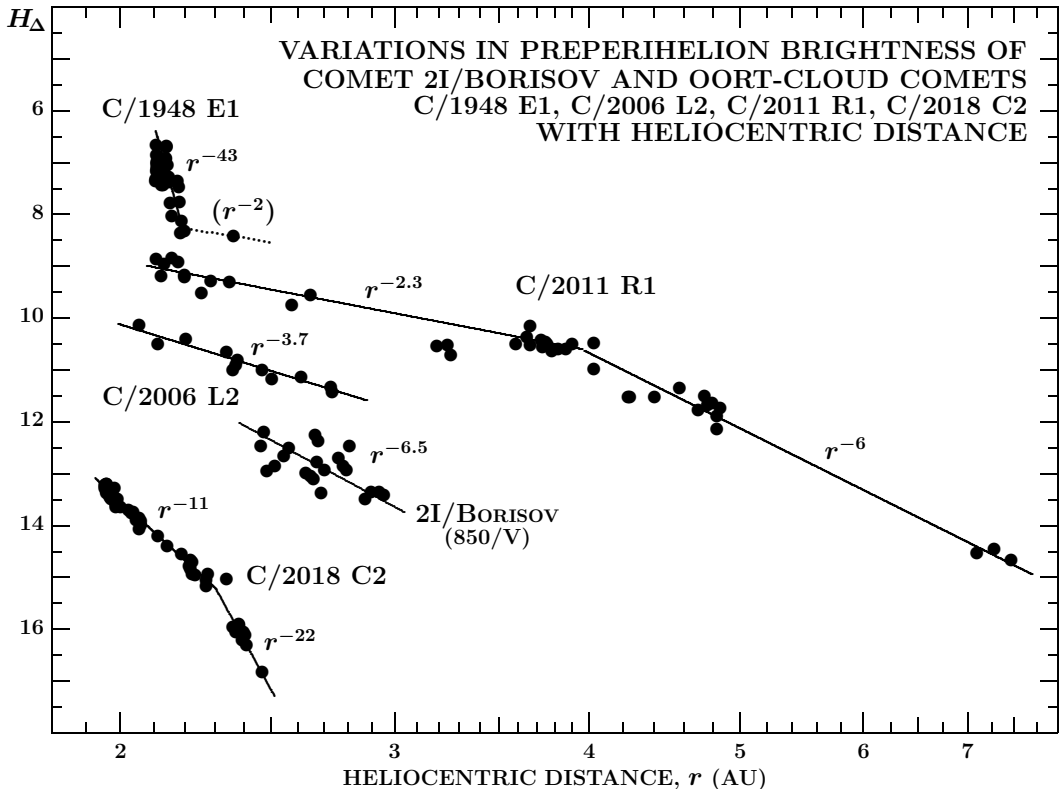}}}
\vspace{-8.4cm}
\caption{Plot of the preperihelion magnitudes from Figure 1 against heliocentric
distance.  It is noted that comet 2I/Borisov now moved up, away from C/2018~C2
and closer to C/2006~L2, an effect that is entirely due to the strongly hyperbolic
orbital velocity of 2I.  The pre-discovery observations of C/2011~R1
are incorporated at the extreme right of the plot.  The curves that cover a wide
enough interval of heliocentric distances consist of two parts, a steeper one
farther from the Sun and a flatter one near perihelion.  Comet C/1948~E1 is an
exception because of a brightness surge it experienced shortly before perihelion.
The future evolution of 2I/Borisov depends on the degree to which the corrected
code 850 data are representative of this comet's total-magnitude
variations.{\vspace{0.75cm}}}
\end{figure*}

It is noted that in contrast to Figure 1, 2I/Borisov is now in a region
closer to the comets of average activity, its brightness varying, if the
converted total visual magnitudes based on the code 850 $V$ data are
representative, as $r^{-6.5}$.  If this fit refers to the pre-break
(large-$r$) law and the break should occur in the coming days, near 2.4~AU
from the Sun, followed by intrinsic variations fitting an $r^{-2}$ rate, 2I
should at perihelion be fainter than normalized magnitude 11 and appear then
close to total visual magnitude 14, unless it undergoes an unexpected
outburst.  On the other hand, if the comet's light curve has since
discovery already been in the post-break segment, i.e., the break
had occurred at a heliocentric distance larger than 3~AU, the comet
should at perihelion be brighter than normalized magnitude 11 and
become more luminous than total apparent visual magnitude 13 when at
its brightest.

Over very short periods of time, on the order of one hour, the apparent
brightness of 2I/Borisov is noticed to vary little, if at all, in some
cases but rather dramatically, by more than 1~mag, in other cases.  Such
peculiar changes could be caused by the object's rotation, but may also
be an instrumental effect caused, for example, by inaccurate centering
of a small scanning aperture on the condensation, especially in the
presence of asymmetric features emanating from it.  To determine the
source of this inconsistency in photometric measurements, I present
in Figure~3 individual overlapping magnitude data by nine different
observers who imaged 2I over the same period of about 70~minutes on September~14.
Four observers --- codes 104, 470, C10, and C95 --- noted no or minor
variations with time, three observers --- codes A77, J95, and Z80 ---
detected general fading, whereas two observers --- codes 970 and B96
--- reported distinct brightening.  The verdict is thus obvious:\ the
trend depends on the observer and is not a product of any physical
changes of the comet.

\section{Dust Tail Orientation and Projected Length}
To build up a dust tail takes time.  The tail's outlines are determined
in part by the ejected grains' velocities, which generally broaden the
feature, but primarily by solar radiation pressure, which for each dust
grain varies as its cross-sectional area and inversely as its mass.  As
a result, the magnitude of the effect varies inversely as the grain's
linear dimension.  In addition, a grain's position in the tail depends
substantially on the time of its ejection.  Solar radiation pressure does
not force any grain out of the comet's orbital plane but makes it move
in a gravitational field that is weaker than the Sun's gravitational field.
The conservation-of-orbital-momentum law requires that the dust tail's
orientation be always constrained to a sector subtended by the
prolonged radius vector {\boldmath $RV$} (antisolar direction) and the
negative orbital-velocity vector {\boldmath $-V_{\bf orb}$}.  The
deviation from the radius vector increases with the time elapsed since
ejection, the dust released very long before perihelion trailing the nucleus
in the orbit along the {\boldmath $-V_{\bf orb}$} vector.

\begin{figure}[t]
\vspace{-5.72cm}
\hspace{2.7cm}
\centerline{
\scalebox{0.92}{
\includegraphics{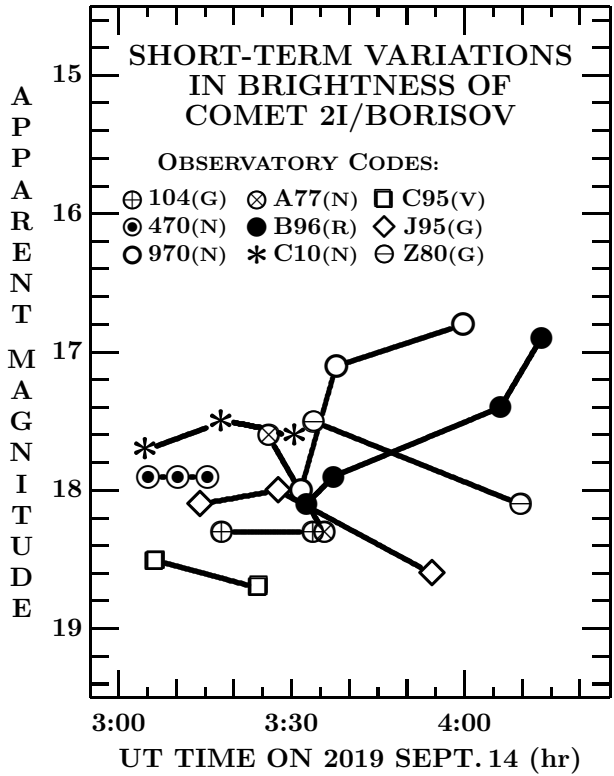}}}
\vspace{-11.4cm}
\caption{Short-term variations in the apparent brightness of 2I/Borisov, measured
in images obtained at nine different observing sites, whose codes and magnitude
types (measured with a variety of scanning apertures and in a variety of
wavelength regions) are shown in the figure.  In the course of the 70-minute
period some sets show little or no variation with time, while other display a
systematic increase or decrease, the evidence that the variations are an
instrumental effect, not intrinsic to the comet.{\vspace{0.75cm}}}
\end{figure}

When a tail is narrow and with no curvature, its direction is measured
with high accuracy, so that comparison is possible with synchrones that
unequivocally determine the times of dust grains' release from the nucleus.
Just as the direction to a single dust grain's location from the nucleus,
the orientation of the tail in projection onto the plane of the sky is defined
by a position angle (reckoned from the north through east), PA$_{\rm tail}$,
that relative to the position angle of the radius vector, PA({\boldmath
$RV$}), and of the negative orbital-velocity vector, PA({\boldmath
$-V_{\bf orb}$}), satisfies a condition
\begin{equation}
{\rm PA}(\mbox{\boldmath $-V_{\bf orb}$}) < {\rm PA}_{\rm tail} < {\rm
 PA}(\mbox{\boldmath $RV$}),
\end{equation}
when the Earth is located above the comet's orbital plane (i.e., the comet
is seen from the Earth to orbit the Sun counterclockwise); or
\begin{equation}
{\rm PA}(\mbox{\boldmath $RV$}) < {\rm PA}_{\rm tail} < {\rm
 PA}(\mbox{\boldmath $-V_{\bf orb}$}),
\end{equation}
when the Earth is below the plane.  It is assumed in these expressions
that \mbox{$|$PA({\boldmath $RV$})\,$-$\,PA({\boldmath
$-V_{\bf orb}$})$| < 180^\circ$}; if this condition is not satisfied,
it is necessary to restore it by adding 360$^\circ$ to the smaller of the
two angles or by subtracting 360$^\circ$ from the larger one.  At the time
of the Earth's transit across the comet's orbital plane \mbox{PA({\boldmath
$-V_{\bf orb}$})\,=\,PA({\boldmath $RV$})}\,or\,\mbox{PA({\boldmath
$-V_{\bf orb}$})\,=\,PA({\boldmath $RV$})$ \pm 180^\circ$}, depending
on the viewing geometry. 

The issues addressed next are (i) the temporal evolution of dust content
in the coma and tail of Oort Cloud comets with perihelia near 2~AU
relative to those permanently beyond the snow line;~and~(ii)~a~modus~operandi
to find out where does 2I/Borisov fit into this scheme.
I first compare images of 2I and two Oort Cloud~analogs in
Figure~4~and~\mbox{Table 2}.\footnote{The images were
selected from the database available at the French website {\tt
http://lesia.obspm.fr/comets/index.php}.}~The~choice~of~the two Oort Cloud
comets reflects the need to examine both a span
of activity and a range of orbital position.  The displayed
images of 2I, C/2014~AA$_{52}$, and C/2011~R1 parallel the differences
in the light curves.  Even though no overall light curve
is available~for~C/2014~AA$_{52}$, the nuclear magnitude reported with the
image~(code~W96) is 16.9 in an aperture of 16\,600 km in diameter, which is
equivalent to a normalized magnitude 14.7 after correcting for the geocentric
distance and phase effect.  A total CCD magnitude measured by code Q62 about
12~days apart~(see~footnote 3) was converted to the time of imaging observation
to suggest the comet's normalized magnitude of 13.2, 1.5~mag brighter.  With
the standard correction to the total visual normalized~magnitude,
C/2014~AA$_{52}$ would be plotted in Figure~2~at~11.7 for \mbox{$r = 2.153$
AU}, 3~mag brighter than the normalized nuclear magnitude. For~C/2011~R1,
the red nuclear magnitude reported~with~the~image~(code~C10)~is~15.7~in~an
aperture of 22\,800~km in diameter, equivalent to a normalized nuclear
magnitude of 13.1.  This implies an expected magnitude
of $\sim$10~in~Figure~1~at~\mbox{$t \!-\! t_\pi \simeq +170$ days},
in perfect agreement with the
fitted light curve.  Finally, the magnitude reported with the image of 2I
(code A77) is 16.8 in an aperture of 24\,200~km in diameter, equivalent to
a normalized nuclear magnitude of 14.1.  One would expect the converted total
visual magnitude of $\sim$11, but the fit in Figure~2 shows the comet
a little more than 1~mag fainter.  2I/Borisov looks much
brighter than C/2014~AA$_{52}$ in Figure~4, due apparently to an instrumental
effect (f/2.8 vs f/8).

\begin{figure*}[ht]
\vspace{-10.2cm}
\hspace{0cm}
\centerline{
\scalebox{0.9}{
\includegraphics{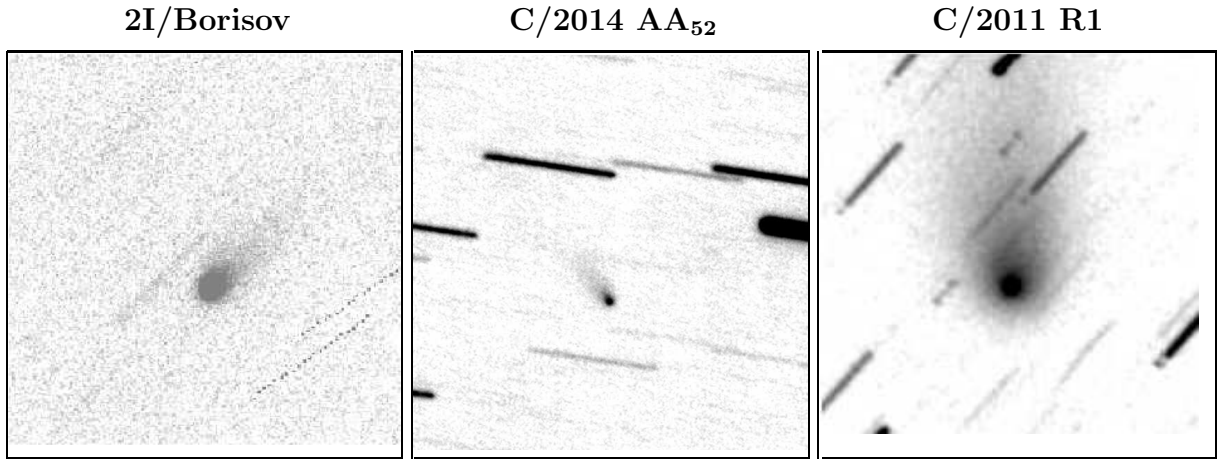}}}

\vspace{-10.13cm}
\caption{Comparison of the appearance of comets 2I/Borisov, C/2014~AA$_{52}$
(Catalina), and C/2011~R1 (McNaught).  The image of 2I was taken by C.~Rinner
and F.~Kugel on 2019 October~8.17~UT with a 40-cm f/2.8 reflector + CCD of the
Observatoire Chante-Perdrix, Dauban, France (code A77); the image of
C/2014~AA$_{52}$ by A.~Maury and J.-F.~Soulier on 2014 December~24.19~UT with
a 40-cm f/8 Ritchey-Chr\'etien reflector + CCD of the Campo Catino Austral
Observatory, San Pedro de Atacama, Chile (code W96); and the image of C/2011~R1
by J.-F.~Soulier on 2013 April~7.06~UT with a 30-cm f/3 Newtonian reflector +
CCD of the Observatoire Maisoncelles, Saint-Martin-du-Boschet, France (code C10).
The common linear scale of the images is 330\,000~km on a side.  The north is
up, east to the left for 2I and C/2014~AA$_{52}$, but the south is up and east
to the right for C/2011~R1. (Image credits:\ F.~Kugel for 2I; J.-F.~Soulier for
C/2014~AA$_{52}$ and C/2011~R1.)}

\vspace{-9.47cm}
\begin{picture}(300,80)
\multiput(30,45)(154,0){3}{\line(1,0){149}}
\multiput(30,-109)(154,0){3}{\line(1,0){149}}
\multiput(29.5,45)(153.5,0){3}{\line(0,-1){154}}
\multiput(179,45)(154,0){3}{\line(0,-1){154}}
\put(105.5,56){\makebox(0,0){\large \bf 2I/Borisov}}
\put(259,56){\makebox(0,0){\large \bf C/2014 AA{\boldmath $_{52}$}}}
\put(412,56){\makebox(0,0){\large \bf C/2011 R1}}
\end{picture}
\vspace{7.4cm}
\end{figure*}

The image of 2I in Figure 4 is reminiscent of a typical Oort Cloud comet at large
heliocentric distance, even more so than the image of C/2011~R1, a comet with
a more sizable nucleus and much more dust in the coma (Section~7).
The position-angle range allowed for the tail by the condition (3) or (4) spans
36$^\circ$ for 2I, 56$^\circ$ for C/2014 AA$_{52}$, and 95$^\circ$ for C/2011~R1.
For all three comets the tail orientation fits the allowed range and Table~2
shows that in each case the tail's position angle, measured by the author from
the images, is closer to the projected {\boldmath $-V_{\bf orb}$} vector than
the {\boldmath $RV$} vector.  Comparison with radiation pressure computations
suggests that dust grains in the tails of 2I and C/2014~AA$_{52}$, both
observed about two months before perihelion, were ejected from the nucleus some
200 days before perihelion, so that the tails were a little less than five
months old when imaged.  Because of the much higher orbital velocity of 2I,
however, the heliocentric distance of this object at ejection was nearly 5~AU,
while C/2014~AA$_{52}$ was only a little over 3~AU from the Sun when dust in its
tail was released.  The dust in the tail in the post-perihelion image of
C/2011~R1 was ejected only about 50~days before perihelion, at a heliocentric
distance of 2.2~AU; the tail thus was more than seven months old.

Insight into the physical processes involved is provided by another image of
C/2011~R1, obtained by C.~Rinner at Dauban on 2012 May~16.95~UT, 156~days before
perihelion and about 100~days before the formation of the~tail seen in the
image taken on 2013 April~7.~The~2012 image shows a tail extending at an
estimated position angle~of 310$^\circ$ and released around 170~days before
Rinner's observation.  This suggests that the tail is being~replenished
as the comet orbits the Sun, but only by millimeter-sized and larger grains
that show up in the tail only after having been subjected to radiation pressure
over some period of time.  This tail should project at a position angle
of 173$^\circ$ in the 2013 image, but is missing.  The obvious conclusion
is that the grains disintegrated between the two observations,
that is, in the general proximity of perihelion, which took place in October 2012.

The implied scenario is strongly reminiscent of comet
C/1980~E1 (old designation 1980~I = 1980b; Bowell), an object with a perihelion
distance of 3.36~AU.  I noted that as the comet was approaching 5~AU on its
way to perihelion, it appeared to display no obvious signs of contemporaneous
activity but clear evidence of powerful emission, deep in the past, of
millimeter-sized and larger dust grains expanding at extremely low,
submeter-per-second, velocities (Sekanina 1982); the heliocentric distance
of the dust emission was estimated at approximately 12~AU.  Subsequently,
A'Hearn et al.\ (1984) reported observations of the OH emission, first detected
at about 5~AU, peaking between 4.5 and 5~AU preperihelion, and then subsiding,
but with brief major outbursts overlapping.  Linking their work to my
finding, A'Hearn et al.\ interpreted OH as a product of the sublimation of
water from the pre-existing halo of grains.  They also remarked that, by
contrast, CN appeared only when the comet was near perihelion.

Returning now to C/2011 R1, the tail that diappeared and was replaced with
a more recent one, is likely to indicate a process experienced by C/1980~E1,
except for taking place at temperatures about 40\,K higher.  If activity of 2I
follows the pattern of Oort Cloud comets, it too should undergo such changes,
although perhaps on a smaller scale than C/2011~R1.  An important, consistent
property of all three comets in Table~2 is the absence of microscopic dust.
In this sense, the behavior of 2I and the Oort Cloud analogs is alike.

Next is the issue of dust grain dimensions in the tail.  Since the smaller the
grain is the higher radiation pressure acceleration it is subjected to, the
projected length of the tail provides, in the synchronic model, information on
grains whose dimensions are the minimum.  Both Oort Cloud comets in Table~2
consistently suggest that the smallest grains in the tail are millimeter sized.
Whether or not the lower limit for 2I is significant remains to be seen.
Unlike the orientation of a tail, its projected length is a very inaccurate
quantity that depends critically on observational conditions, instrumentation,
exposure time, etc.

Data for a second test of the tail orientation and length have been provided
by Beyer's (1950) visual monitoring of comet C/1948~E1 over a period of many
months, as already remarked on briefly in Section~4.  Figure~5 exhibits his
measurements of temporal variations in the position angle of the tail; they
are compared with the position angles of the {\boldmath $RV$} and {\boldmath
$-V_{\bf orb}$} vectors as well as with those of a few synchrones.  It is
noted that only one out of the more than 50 observations marginally defies the
condition of the tail's confinement to the sector subtended by the two vectors.
This remarkable agreement is as much a testimonial to the exceptionally high
quality of Beyer's observational work as it is a confirmation that the model
of dust tails works.\footnote{The measurements that Beyer marked as being
uncertain have not been plotted.}

\begin{table*}[t]
\vspace{-4.18cm}
\hspace{-0.5cm}
\centerline{
\scalebox{1}{
\includegraphics{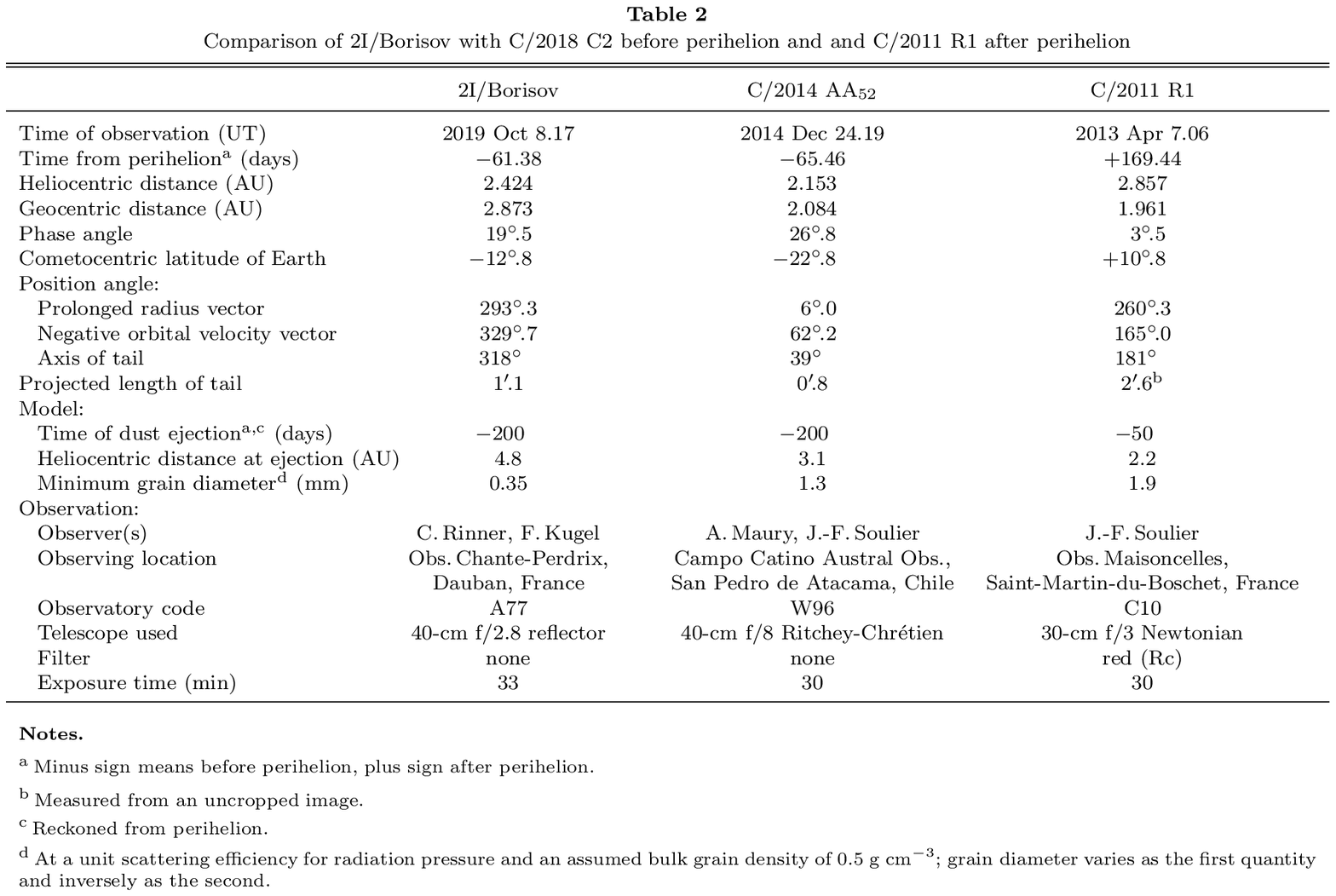}}}
\vspace{-12.5cm}
\end{table*}

The diagnostic part of Figure 5 is the post-perihelion period
of time when the difference between
the position angles of the vectors {\boldmath $RV$} and {\boldmath
$-V$} increases toward the maximum of 180$^\circ$ at the time of
the Earth's transit across the comet's orbital plane, several weeks
after Beyer's measurements of the tail terminated.

In the period of time between 50 and 150 days after perihelion
the tail's position angle stays close to the position angle of
the negative orbital velocity vector and varies in a manner that
parallels the variations in the latter, but is entirely dissociated
from the prolonged radius vector.  The synchrones plotted in
Figure~5 suggest that the observed tail consisted primarily of
dust ejected between 300 and 100 days before perihelion.  Beyer
(1950) stated that the tail was slender, which suggests that no
dust related to the post-perihelion activity was detected by his
tail observations.

Figure 6 shows a plot of Beyer's (1950) measurements of the projected length
of the tail of C/1948~E1.  Considerable scatter is now obvious, unquestionably
because of variable seeing and the comet's gradual fading.  It turns out that
around 130~days after perihelion Beyer could see only the part of the tail
containing grains exceeding 1~mm in diameter, six times as large as those he
observed when the comet was near perihelion.

Given the in-depth description of the complex evolution of the dust tails of Oort
Cloud comets with perihelia near 2~AU and in view of the early signs that the tail
of 2I/Borisov has been developing in a similar manner, I consider it useful to
provide an extensive tail ephemeris for the interstellar object to assist in the
forthcoming investigations of this type.  Presented in Table~3, the individual
columns list the ephemeris date (at a 20-day step), the solar elongation,
the position vectors of the radius vector and negative orbital-velocity
vector, the cometocentric latitude of the Earth, and, for each of three ejection
times --- 500 and 200 days before perihelion and at perihelion --- the predicted
position angle of the tail axis and the ``extent,'' which is the projected length
of the tail defined by the location~of dust grains subjected to a radiation
pressure acceleration of 0.001 the Sun's gravitational acceleration; if the
scattering efficiency for radiation pressure of such grains is unity and their
bulk density 0.5~g~cm$^{-3}$, they are 2.3~mm~in~diameter, the smallest in the
tail.  For
a given radiation pressure acceleration the grain size varies as the scattering
efficiency and inversely as the bulk density.  From the position angle of the
tail measured in an image, the ejection time is determined by interpolating the
tabulated position angles, then from the tail's measured projected length the
maximum radiation pressure acceleration is determined as 0.001 times the ratio of
the projected length (in minutes of arc) to the interpolated value of the tabulated
extent. For a given scattering efficiency and bulk density, the minimum diameter of
grains in the tail varies inversely as the radiation pressure acceleration.

\begin{figure}[ht]
\vspace{-1.1cm}
\hspace{2.8cm}
\centerline{
\scalebox{0.85}{
\includegraphics{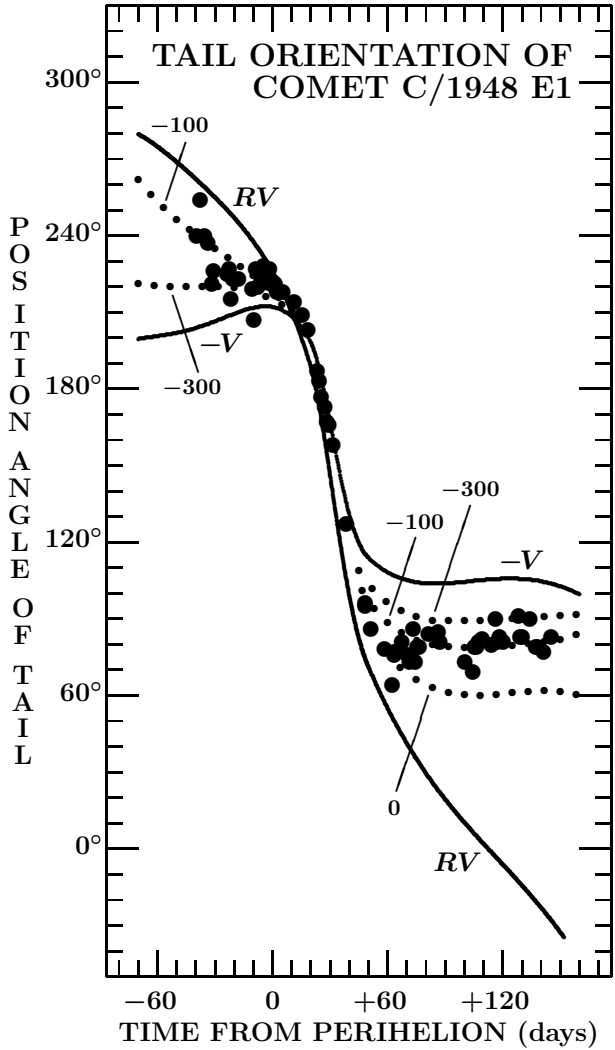}}}
\vspace{-10.45cm}
\caption{Plot of Beyer's (1950) measurements of the position angle of the
tail axis of C/1948~E1 as a function of time reckoned from perihelion. The
curves of temporal variations in the radius vector and the negative
orbital-velocity vector are marked by {\boldmath $RV$} and {\boldmath
$-V_{\bf orb}$}, respectively.  The dotted curves are position angles for
ejection times 300 and 100~days before perihelion and at perihelion (0).  Note
that the observations made more than 50~days after perihelion indicate
the tail was consistently oriented much closer to the direction of the
{\boldmath $-V_{\bf orb}$} vector than the {\boldmath $RV$} vector,
confirming the absence of new emissions of microscopic dust.  The comet
passed through perihelion on 1948 May 16.61 TT.{\vspace{0.5cm}}}
\end{figure}
\begin{figure}[t]
\vspace{-4.06cm}
\hspace{2.3cm}
\centerline{
\scalebox{0.85}{
\includegraphics{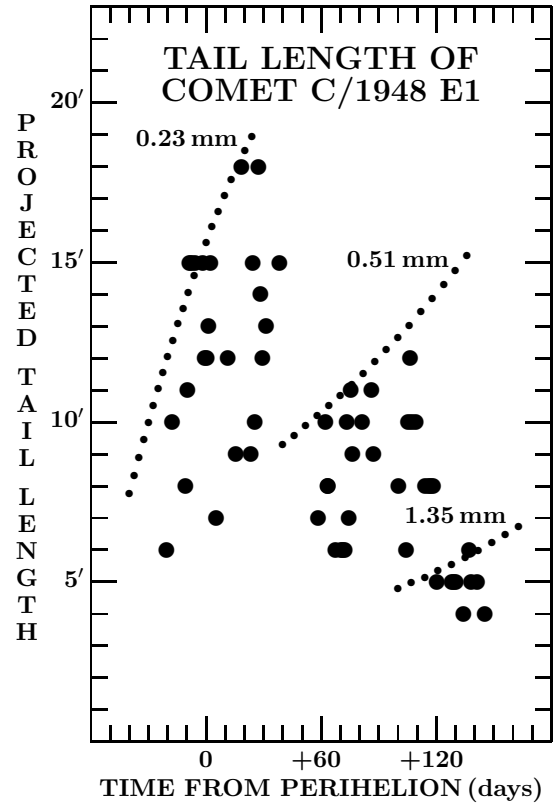}}}
\vspace{-10.75cm}
\caption{Plot of Beyer's (1950) measurements of the projected length of
the tail of C/1948~E1 as a function of time reckoned from perihelion.  The
dotted curves are the temporal variations in the projected distance from the
nucleus of dust grains ejected 200 days before perihelion and subjected to
solar radiation pressure accelerations of, respectively, 0.010 (the steepest
curve), 0.0045, and 0.0017 (the least steep curve) the Sun's gravitational
acceleration; at a unit scattering efficiency for radiation pressure and an
assumed bulk grain density of 0.5~g~cm$^{-3}$ the smallest grains detected
by Beyer around perihelion, near 80 days after perihelion, and close to
130 days after perihelion are, respectively, 0.23~mm, 0.51~mm, and 1.35~mm
in diameter, marking the dotted curves.{\vspace{0.6cm}}}
\end{figure}

\section{Dust Production Along the Preperihelion Leg of the Orbit}
In order to compare the amount of dust in the coma of a comet
from non-uniform measurements, A'Hearn et al.\ (1984) introduced a
product of an albedo, filling factor of dust in an aperture, and
the aperture's size, {\it Af}$\rho$, a parameter that can easily
be measured.  Both A'Hearn et al.\ and Fink \& Rubin (2012), who
investigated the physical meaning of the parameter in considerable
detail, focused on the {\it Af}$\rho$ parameter as a proxy of the dust
production rate.  This implied relationship is justified on the grounds
that the filling factor $f$ depends on the number of dust grains
in the column within the aperture, which in turn is proportional
to the ratio of the dust production rate and dust expansion velocity.
Fink \& Rubin concluded that submicron-sized grains, whose velocities
are in the subkilometer-per-second range, contribute a large fraction
to {\it Af}$\rho$.  With these velocities, grains should evacuate the
circumnuclear volume of space several tens of thousands kilometers in
radius in a matter of one day or less, so this scenario naturally
requires that new ejecta replace the lost dust.  As the dust production
rate is expected to correlate with the light curve, it should vary {\it
inversely\/} as a power of heliocentric distance (Section~5).

\begin{table*}[t]
\vspace{-4.18cm}
\hspace{-0.5cm}
\centerline{
\scalebox{1}{
\includegraphics{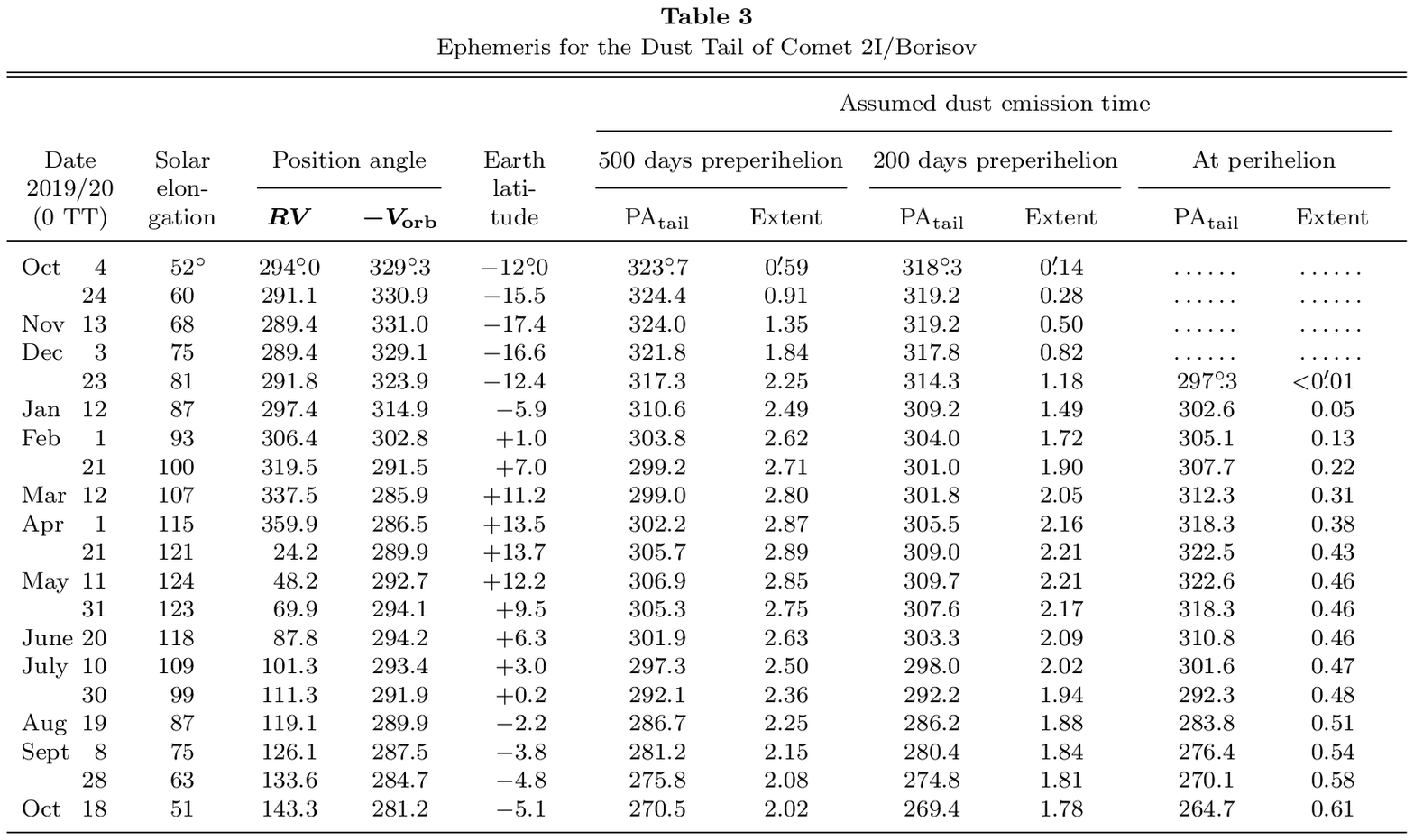}}}
\vspace{-14.8cm}
\end{table*}

The evidence from Section 6 overwhelmingly indicates that the tails
of the 2I's analogs arriving from the Oort Cloud contained old,
millimeter-sized and larger grains moving --- as the tails' narrow
breadth implies --- at very low velocities.  Because the dust in the tail
is near the lower end of the grain population's size spectrum (i.e., near
the upper end of the range of radiation pressure accelerations), still
larger grains moving at still lower velocities should occupy the coma
probed by the {\it Af}$\rho$~parameter, some perhaps even gravitationally bound
to the nucleus (at least along an early part of the inbound leg of the orbit).
Under these circumstances, the {\it Af}$\rho$ parameter measures the coma's
(surviving) dust content along with any potential recent dust production rate.

With these arguments in mind, I inspected the {\it Af}$\rho$ database
at the website of the Spanish Comet Hunters,\footnote{See {\tt
http://astrosurf.com/cometas-obs}; the website is maintained by J.~Castellano,
E.~Reina, and R.~Naves.} in which all entries are provided for
a normalized aperture radius of 10\,000~km.  For correlation with
Figure~2, I was interested only in the {\it preperihelion\/} data.
Fairly sizable sets exist for C/2006~L2 (23~data points), C/2011~R1
(10), C/2014~AA$_{52}$ (20), and C/2018~C2 (12).  There are 12~data
points available for 2I, spanning the period of September~9 through
October~8.

\begin{figure}[b]
\vspace{-4.2cm}
\hspace{1.65cm}
\centerline{
\scalebox{0.72}{
\includegraphics{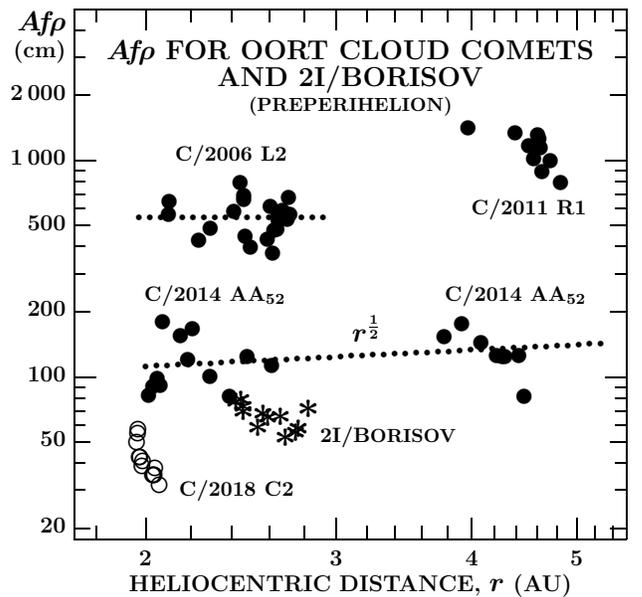}}}
\vspace{-9.05cm}
\caption{Plot of the dust production proxy parameter {\it Af}$\rho$ as
a function of heliocentric distance $r$.  The data, normalized to a zero
phase angle, are preperihelion measurements from a number of observatories
reporting to the Spanish website of Comet Hunters, referring to four Oort
Cloud comets --- C/2006~L2, C/2011~R1, C/2014~AA$_{52}$, C/2018~C2 ---
and to 2I/Borisov.  The {\it Af}$\rho$ parameter decreases with
decreasing $r$ for C/2014~AA$_{52}$ and stays essentially constant over
a shorter interval of $r$ for C/2006~L2.  For C/2011~R1 and C/2018~C2 the
range of $r$ covered by the observations is too short to detect a trend;
this is also true for 2I.{\vspace{-0.05cm}}}
\end{figure}

I corrected the {\it Af}$\rho$ data for the phase effect with the use
of the Marcus (2007) law and then plotted them against heliocentric
distance $r$ in Figure~7.  The particularly striking feature is the plot
for C/2014~AA$_{52}$, the only comet for which the data cover a very wide
range of $r$:\ the {\it Af}$\rho$ value {\it decreases\/} with decreasing
{\vspace{-0.06cm}}heliocentric distance as $r^{\frac{1}{2}}$!  Although the
light curve of this comet is not well established, five measurements of the
total CCD magnitude by a single observer (code Q62) suggest that
between 2.2~AU and 4.9~AU the preperihelion intrinsic brightness
varied as approximately $r^{-3}$, a rate about midway between
C/2011~R1 and C/2006~L2.  In any case, the preperihelion light curve
and {\it Af}$\rho$ are absolutely incompatible.  For C/2006~L2
the range of heliocentric distances covered by the preperihelion
measurements of {\it Af}$\rho$ is shorter, yet wide enough to show
essentially no change between 2.0~AU and 2.7~AU from the Sun.  For
2I/Borisov the same result is suggested by the measurements between
2.4~AU and 2.9~AU from the Sun.  The preperihelion intervals of $r$ for
C/2011~R1 and C/2018~C2 are too short to allow a conclusion on the trend.
However, the post-perihelion {\it Af}$\rho$ data available for C/2011~R1 are
lower than the preperihelion ones and dropping as an inverse 4.5 power of
heliocentric distance.  The overall trend of {\it Af}$\rho$ is reminiscent
of the variations observed in C/1980~E1 by A'Hearn et al.\ (1984), who
determined that this parameter's value dropped from \mbox{7000--8000 cm}
near 5~AU preperihelion to \mbox{3000--4000 cm} near perihelion, thus
following an $\sim \! r^2$ law, much steeper than C/2014~AA$_{52}$ in Figure~7.

Among the Oort Cloud comets with perihelia near 2~AU, {\it Af}$\rho$ appears to
correlate with the light curve only in the sense that the brighter the comet is
intrinsically at perihelion, the higher average value of {\it Af}$\rho$ it has.
Comet 2I/Borisov seems so far to fit this relationship --- at least to the extent
that one can find out from the limited database --- on the fainter side; future
observations should show whether this preliminary conclusion will continue to hold.

\section{Discussion and Conclusions}
In summary, the presented evidence strongly suggests that the pattern
of activity of Oort Cloud comets with perihelia near 2~AU differs substantially
from that of Oort Cloud comets whose perihelia exceed 4~AU.  The difference
stems from the position of the snow line in the Solar System --- the
more distant comets never experience the nucleus activity driven by
the sublimation of water ice.  On the other hand, both categories display
effects of the absence of microscopic grains in the dust ejecta.

The snow line has ramifications for the evolution of the halo
of large grains in the atmospheres of Oort Cloud comets.  For example, the
sublimation lifetime of an isothermal, pure icy grain of a bulk
density of 0.5~g~cm$^{-3}$ and initially 1~mm in diameter amounts to
more than 4~millennia at a heliocentric distance of 5~AU, 23~years
at 4~AU, 85~days at 3.2~AU, 2~days at 2.5~AU, and about 8~hours
at 2~AU.  However, these numbers are only crude lower limits, because
actual grains are not pure ice, but a mixture of refractory material
and ice.  If ice serves as a glue that holds the grains
together, its gradual evacuation should progressively weaken the
grains' cohesion and result eventually --- over much longer periods of
time than are the above lifetimes --- in their disintegration.
And as this process is strongly heliocentric-distance dependent, the
tail has an increasing tendency to disappear and the
{\it Af}$\rho$ value of the coma to drop steadily as the heliocentric
distance continues to decrease below 4~AU.  In addition, this process
is accelerated by the fact that as the grains become less massive, the
radiation pressure effects on them increase, thus speeding up their
dispersion in space, away from the comet.

The absence of microscopic grains in the tails of Oort Cloud comets
(their existence in the coma but not in the tail would require extremely
short lifetimes) readily explains
the instances of a missing dust tail near the radius vector.  Sizable
grains, subjected to radiation pressure accelerations not exceeding
1~percent of the Sun's gravitational acceleration, reach the tail only
after a prolonged time (if they survive), drifting in the meantime through
the coma following their low-velocity release from the nucleus.  By
the time they get into the tail, their trajectories deviate notably
from the direction of the radius vector.

As I concluded more than 40 years ago (Sekanina 1975), preperihelion
activity of Oort Cloud comets is essentially continuous,
with possible occasional surges, and those with perihelia below $\sim$4~AU
undergo nucleus activity that is driven by the sublimation of water ice.
Products of this activity overlap the surviving part of the original halo
of millimeter-sized and larger grains.  This scenario is consistent with
A'Hearn et al.'s (1984) extensive observations of C/1980~E1, whose
perihelion distance was 3.36~AU, as well as with overwhelming evidence
of ``tail rejuvenation'' among the comets with perihelia near 2~AU
examined in this paper.

The complex evolution of Oort Cloud comets with perihelia at this distance
from the Sun is illustrated by C/2011~R1.  Its high preperihelion values
of the {\it Af}$\rho$ parameter imply the presence of a halo of early,
nearly stationary dust grains in the atmosphere at heliocentric distances
between 4 and 5~AU from the Sun.  A preperihelion image shows a tail
consisting of grains released at about 4~AU from the Sun, while the
comet's post-perihelion image in Figure~4 (and described in Table~2)
documents the presence of dust grains whose ejection from the comet
dates back to 2~months or so before perihelion.  Thus, the products of
the early activity failed to survive past perihelion.  This finding
is consistent with the evolution of another Oort Cloud comet, C/1950~K1
(Minkowski), of perihelion distance 2.57~AU, which I examined closely
earlier (Sekanina 1975).  The tail directions measured in some 40~images,
taken over a period of nearly 300~days, showed that the observed dust ejecta
were nearly always released from the comet between $\sim$50 and $\sim$300
days before observation and as late as the perihelion time.  By contrast,
Oort Cloud comets with perihelia near or beyond the snow line display
evidence of early preperihelion activity at any subsequent time,
including long after perihelion.  For example, my detailed analysis
of C/1954~O2 in an image taken by Osterbrock (1958) about 100~days
after perihelion, when the comet was 3.96~AU from the Sun, shows that
the outer extremity of the tail was populated by grains $\sim$2~mm in
diameter released from the nucleus nearly three years before perihelion,
at a heliocentric distance of 8.8~AU and that minor contributions to
the tail may have been provided by grains ejected as late as 200~days
before perihelion, 4.2~AU from the Sun, but not later (Sekanina 1975).

Accordingly, evidence shows that dust activity of most, if not all, Oort
Cloud comets with perihelion distances under consideration drops rather
rapidly, or possibly ceases for some, after
perihelion; this does by no means prevent these comets from continuing to
outgas.  Among the objects examined in this paper, neither C/2011~R1
(Table~2) nor C/1948~E1 (Figure~5) show evidence for post-perihelion dust
ejecta in their tails in images taken many months after perihelion.  Also
supporting this notion, though less conclusively, are (i)~the
systematic post-perihelion drop in the values of {\it Af}$\rho$ of all
comets in Table~1, for which the data are available; and (ii)~the
post-perihelion fading that is steeper than the preperihelion
brightening seen in the light curves (Whipple 1978 and Figure~1).

Judging from the six weeks of data, the interstellar comet 2I/Borisov fits,
in broad terms, the range of properties of Oort Cloud comets of the
same perihelion distance as to its appearance, tail shape and orientation,
and the dust production proxy parameter {\it Af}$\rho$.  The available
values of {\it Af}$\rho$ suggest that the comet is a relatively
small object, with probably a subkilometer-sized nucleus.  Yet, the comet's
motion may be subjected,~if at all, to only minor nongravitational effects.
The~detections of CN near 2.7~AU (Fitzsimmons et al.\ 2019a) and C$_2$ at
2.5~AU from the Sun (Kareta et al.\ 2019; although questioned by Fitzsimmons
et al.\ 2019b) also fit the general behavior of these Oort Cloud comets.  The
implied carbon chain depletion in 2I (regardless~of whether C$_2$ was or was not
detected) compares favorably with evidence on C/1999~S4 (Farnham et al.\
2001), a probable Oort Cloud comet.  Comet Borisov's
activity appears to have already been driven by the sublimation of water
ice in the past weeks; whether water is going to be
detected is a matter of the emission rate and instrumental sensitivity.
The comet's observations near and after perihelion will allow examination
of a variety of diagnostic activity signatures to test the degree of 2I's
similarity to Oort Cloud comets.  The predicted tail ephemeris in
Table~3 is intended to aid investigations of the evolution of dust production
as the comet continues to pursue its motion along the strongly hyperbolic orbit.\\
 
This research was carried out at the Jet Propulsion Laboratory, California
Institute of Technology, under contract with the National Aeronautics and
Space~Administration.\\[-0.1cm]
\begin{center}
{\footnotesize REFERENCES}
\end{center}
\vspace{-0.4cm}
\begin{description}
{\footnotesize
\item[\hspace{-0.3cm}]
A'Hearn, M.\ F., Schleicher, D.\ G., Feldman, P.\ D., et al.\ 1984, AJ,{\linebreak}
 {\hspace*{-0.6cm}}89, 579
\\[-0.57cm]
\item[\hspace{-0.3cm}]
Bassot, J.\ A.\ L.\ 1907, Astron.\ Nachr., 174, 127
\\[-0.57cm]
\item[\hspace{-0.3cm}]
Beyer, M.\ 1950, Astron.\ Nachr., 278, 217
\\[-0.57cm]
\item[\hspace{-0.3cm}]
Bolin, B., Denneau, L., Wainscoat, R., et al.\ 2014, CBET 3812
\\[-0.57cm]
\item[\hspace{-0.3cm}]
Dehnen, W., \& Binney, J.\ J.\ 1998, MNRAS, 298, 387
\\[-0.57cm]
\item[\hspace{-0.3cm}]
de Le\'on, J., Licandro, J., Serra-Ricart, M., et al.\ 2019, Res.\,Notes{\linebreak}
 {\hspace*{-0.6cm}}AAS, 3 (9), 131
\\[-0.57cm]
\item[\hspace{-0.3cm}]
Dybczy\'nski, P.\ A., Kr\'olikowska, M., \& Wysocza\'nska, R.\ 2019,{\linebreak}
 {\hspace*{-0.6cm}}eprint arXiv:1909.10952
\\[-0.57cm]
\item[\hspace{-0.3cm}]
Elenin, L.\ 2012, MPC 78724 and 78753
\\[-0.57cm]
\item[\hspace{-0.3cm}]
Farnham, T.\ L., Schleicher, D.\ G., Woodney, L.\ M., et al.\ 2001,{\linebreak}
 {\hspace*{-0.6cm}}Science, 292, 1348
\\[-0.57cm]
\item[\hspace{-0.3cm}]
Fink, U., \& Rubin, M.\ 2012, Icarus, 221, 721
\\[-0.64cm]
\item[\hspace{-0.3cm}]
Fitzsimmons, A., Hainaut, O., Yang, B., et al.\ 2019a, CBET 4670}
\\[-0.57cm]
\end{description}
\vspace{9.5cm}
\pagebreak
\begin{description}
{\footnotesize
\item[\hspace{-0.3cm}]
Fitzsimmons, A., Opitom, C., Hainaut, O., et al.\ 2019b, CBET{\linebreak}
 {\hspace*{-0.6cm}}4679
\\[-0.57cm]
\item[\hspace{-0.3cm}]
Gibbs., A.\ R., Grauer, A.\ D., Leonard, G.\ J., et al.\ 2018, MPC{\linebreak}
 {\hspace*{-0.6cm}}109188 and 109209
\\[-0.57cm]
\item[\hspace{-0.3cm}]
Green, D.\ W.\ E.\ 2019, CBET 4666
\\[-0.57cm]
\item[\hspace{-0.3cm}]
Guzik, P., Drahus, M., Rusek, K., Waniak, W., et al.\ 2019, eprint{\linebreak}
 {\hspace*{-0.6cm}}arXiv:1909.05851
\\[-0.57cm]
\item[\hspace{-0.3cm}]
Harrington, R.\ S.\ 1952, PASP, 64, 275
\\[-0.57cm]
\item[\hspace{-0.3cm}]
Hill, R.\ E.\ 2008, IAUC 8945
\\[-0.57cm]
\item[\hspace{-0.3cm}]
Hill, R.\ E., Gibbs, A.\ R., Kowalski, R.\ A., et al.\ 2012, MPC 78723{\linebreak}
 {\hspace*{-0.6cm}}and 78753
\\[-0.57cm]
\item[\hspace{-0.3cm}]
Jewitt, D., \& Luu, J.\ 2019, eprint arXiv:1910.02547
\\[-0.57cm]
\item[\hspace{-0.3cm}]
Kareta, T., Andrews, J., Noonan, J.\ W., et al.\ 2019, eprint arXiv:{\linebreak}
 {\hspace*{-0.6cm}}1910.03222
\\[-0.57cm]
\item[\hspace{-0.3cm}]
Kowalski, R.\ A.\ 2014, CBET 3812
\\[-0.57cm]
\item[\hspace{-0.3cm}]
Kr\'olikowska, M., \& Dybczy\.nski, P.\ A.\ 2013, MNRAS, 435, 440
\\[-0.57cm]
\item[\hspace{-0.3cm}]
Kr\'olikowska, M., Sitarski, G., Pittich, E.\ M., et al.\ 2014, A\&A,{\linebreak}
 {\hspace*{-0.6cm}}571, A63
\\[-0.57cm]
\item[\hspace{-0.3cm}]
Marcus, J.\ N.\ 2007, Int.\ Comet Quart., 29, 39
\\[-0.57cm]
\item[\hspace{-0.3cm}]
Marsden, B.\ G., Sekanina, Z., \& Everhart, E.\ 1978, AJ, 83, 64
\\[-0.57cm]
\item[\hspace{-0.3cm}]
McNaught, R.\ H.\ 2006, IAUC 8721
\\[-0.57cm]
\item[\hspace{-0.3cm}]
McNaught, R.\ H.\ 2011a, IAUC 9213
\\[-0.57cm]
\item[\hspace{-0.3cm}]
McNaught, R.\ H.\ 2011b, CBET 2810 and IAUC 9230
\\[-0.57cm]
\item[\hspace{-0.3cm}]
McNaught, R.\ H.\ 2013, MPC 81939 and 81943
\\[-0.57cm]
\item[\hspace{-0.3cm}]
Meech,\,K.\,J.,\,Pittichov\'a,\,J.,\,Bar-Nun,\,A.,\,et\,al.\,2009,\,Icarus,\,201,\,719
% {\hspace*{-0.6cm}}719
\\[-0.57cm]
\item[\hspace{-0.3cm}]
Merton, G.\ 1949, MNRAS, 109, 248
\\[-0.57cm]
\item[\hspace{-0.3cm}]
Micheli, M.\ 2018, CBET 4501
\\[-0.57cm]
%
% \item[\hspace{-0.3cm}]
% Minor Planet Center 2019, MPEC 2019-S72
% \\[-0.57cm]
%
\item[\hspace{-0.3cm}]
Mount Lemmon Survey 2018, CBET 4501
\\[-0.57cm]
\item[\hspace{-0.3cm}]
O'Callaghan, J.\ 2019, Sci.\ Amer., newsletter dated September 13
\\[-0.57cm]
\item[\hspace{-0.3cm}]
Osterbrock, D.\ E.\ 1958, ApJ, 128, 95
\\[-0.57cm]
\item[\hspace{-0.3cm}]
Porter, J.\ G.\ 1955, MNRAS, 115, 190
\\[-0.57cm]
%
% \item[\hspace{-0.3cm}]
% Marsden, B.\ G., Sekanina, Z., \& Yeomans, D.\ K.\ 1973, AJ,~78,~211
% \\[-0.57cm]
%
\item[\hspace{-0.3cm}]
Roemer, E.\ 1962, PASP, 74, 351
\\[-0.57cm]
\item[\hspace{-0.3cm}]
Sekanina, Z.\ 1973, Astrophys.\ Lett., 14, 175
\\[-0.57cm]
\item[\hspace{-0.3cm}]
Sekanina, Z.\ 1975, Icarus, 25, 218
\\[-0.57cm]
\item[\hspace{-0.3cm}]
Sekanina, Z.\ 1982, AJ, 87, 161
\\[-0.57cm]
\item[\hspace{-0.3cm}]
Sekanina, Z.\ 2017, eprint arXiv:1712.03197
\\[-0.57cm]
\item[\hspace{-0.3cm}]
Sekanina, Z.\ 2019a, eprint arXiv:1901.08704
\\[-0.57cm]
\item[\hspace{-0.3cm}]
Sekanina, Z.\ 2019b, eprint arXiv:1903.06300
\\[-0.57cm]
\item[\hspace{-0.3cm}]
Van Biesbroeck, G.\ 1950, AJ, 55, 53
\\[-0.57cm]
\item[\hspace{-0.3cm}]
Weiss, E.\ 1907a, Astron.\ Nachr., 176, 299
\\[-0.57cm]
\item[\hspace{-0.3cm}]
Weiss, E.\ 1907b, Astron.\ Nachr., 176, 327
\\[-0.57cm]
\item[\hspace{-0.3cm}]
Whipple, F.\ L.\ 1978, Moon \& Plan., 18, 343
\\[-0.57cm]
\item[\hspace{-0.3cm}]
Williams, G.\ V.\ 2014, MPEC 2014-D45
\\[-0.57cm]
\item[\hspace{-0.3cm}]
Wolf, M.\ 1907, Astron.\ Nachr., 176, 315
\\[-0.65cm]
\item[\hspace{-0.3cm}]
Yang, B., Keane, J.\ V., Kelley, M.\ S.\ P., et al.\ 2019, CBET 4672}
\vspace{9.24cm}
\end{description}
\end{document}